\documentclass[aps,eqsecnum,showpacs]{revtex4}

\usepackage{amsmath,graphicx,amssymb}

\newcommand{\diver}{{\rm div}}

\begin{document}

\title{Magnetic Soret effect: application of the ferrofluid dynamics theory}
\author{Adrian Lange}
\affiliation{MPI f\"ur Physik komplexer Systeme,
N\"othnitzer Stra{\ss}e 38,
D-01187 Dresden, Germany}

\date{\today}

\begin{abstract}
The ferrofluid dynamics theory is applied to thermodiffusive problems in
magnetic fluids in the presence of magnetic fields. The analytical form for
the magnetic part of the chemical potential and the most general expression of the
mass flux are given. By employing these results
to experiments, global Soret coefficients in agreement with measurements
are determined. Also an estimate for a hitherto unknown transport coefficient
is made.
\end{abstract}

\maketitle

\section{Introduction}
Magnetic fluids (MFs) are colloidal suspensions of
ferromagnetic nanoparticles dispersed in a nonmagnetic carrier liquid.
MFs behave superparamagnetically in a magnetic field and have a far
reaching application potential spanning
from sealants in rotary shaft to heat dissipaters in loud
speaker coils \cite{handbook} to carrier liquids for medical substances
\cite{clinical_appl99}. Starting in the mid-sixties of the last century,
when MFs were first available, research on those fluids had been proceeding
on the calm fairway of an established and well founded field of research.
Particularly the theoretical work had been based on the achievements
of the two pioneers, Rosensweig \cite{rosensweig} and Shliomis
\cite{shliomis71,shliomis74}. But 10 years ago a series of papers
\cite{liu93,liu95,liu98,mueller01} started to appear pointing
specifically to the deficiencies of the microscopic approach in
\cite{shliomis71,shliomis74} and proposing
`` ... a general, strictly macroscopic approach relying solely on symmetry
considerations, conservation laws, and thermodynamics.''
\cite{mueller_reply03}. This approach, called ferrofluid dynamics (FFD), sparked
an impassioned discussion \cite{shliomis_comment03,mueller_reply03} about which
theory explains better the experimental facts for the reduced viscosity of a MF in an
ac magnetic field \cite{bacri95} or for the magnetovortical resonance
\cite{gazeau97,gazeau96}. That new theory also triggered an experiment
\cite{odenbach02_flow} confirming a proposed nonzero transport coefficient which is
zero in the microscopic approach \cite{shliomis71,shliomis74}. Other
proposed effects as shear-excited sound
\cite{mueller02_sound,mueller_erratum03} await their confirmation yet.

For the description of thermal convection in magnetic fluid, the
fluid has been considered as a one-component fluid with effective properties
in many studies (see \cite{auernhammer00,huang98_thermo,recktenwald98} and references therein).
The limits of this coarse grained view onto the colloidal suspension of
ferromagnetic nanoparticles are just being revealed. Considering a magnetic fluid
as a binary liquid, the thermal convection is found to set in at Rayleigh numbers
well below the threshold for a MF considered as a single-component fluid
\cite{ryskin03}.

The thermodiffusive or Soret effect describes the establishment of concentration
gradients in response to temperature gradients for a two-(or multi-)component
fluid. Since the motion of the ferromagnetic nanoparticles
in the MF can be influenced by external magnetic fields, the Soret effect in
MFs shows a strong dependence on any nonzero magnetic field strength
\cite{voelker02_phdBook,voelker03,voelker03_prl}.
In a vertical layer the Soret coefficient $S_T$ depends {\it non}monotonously
on the strength of the field in the cases where the field is either parallel or
perpendicular to the temperature gradient \cite{voelker03,voelker02_phdBook}.
Contrary, for both orientations of the magnetic field the Soret coefficient depends
monotonously on the strength of the field if the layer is horizontal \cite{voelker03_prl}.
The changes of  $S_T$ can be up to six times its zero field value \cite{voelker03}
and even a change of the sign of $S_T$ was
measured for strong fields \cite{voelker02_phdBook,voelker03,voelker03_prl}.

The known theoretical approaches for the Soret effect in magnetic fluids
\cite{blums98,shliomis02} need as an essential input an expression for the
magnetophoretic velocity of the nanoparticles with respect to the carrier
liquid. For that purpose certain microscopic properties are assumed as
a dilute colloid containing spherical particles of equal size
and the applicability of the Stokes hydrodynamic drag \cite{blums98,shliomis02}.
Also assumptions about the deformation of the temperature distribution around
the particle are made if its thermal conductivity is different from that
of the surrounding carrier liquid \cite{blums98}. A comparison with the
known experimental results shows great differences: the microscopic
theory \cite{blums98,blums_book} gives only changes of $S_T$ which are about {\it three}
orders of magnitude smaller than the experimentally measured ones
(\cite{voelker03} and Fig. 23 in \cite{voelker02_phdBook}). Also in the frame of
a thermodynamic approach \cite{bacri95_forced} it is not possible to describe the drastic
changes of $S_T$ measured in the experiment. That means that
with respect to thermodiffusive processes in MF in the presence of magnetic fields
a wide gap between experiment and theory has to be bridged. Therefore it is the
aim of this work to present a different approach, where in the frame of a
macroscopic theory, the FFD, the experimental results can be described
significantly better.

Usually an external temperature gradient causes both convection and
thermodiffusion in any colloidal suspension. How these two effects are
interacting with each other is not yet finally resolved as the discussion
about the possibility of a state of relaxation-oscillation convection
highlights \cite{shliomis00,ryskin03}. The mutual interference
of convection and thermodiffusion is even more severe if additionally an
external magnetic field is applied as in the case of MFs
\cite{voelker02_phdBook,voelker03,voelker03_prl,shliomis02,shliomis_jmmm02}.
The problems caused by that mutual interference for the determination of
the Soret effect are outlined in \cite{voelker03_prl} and result in a new
experimental setup for a horizontal layer of MF which is analyzed
theoretically in this work.

\section{Ferrofluid dynamics: chemical potential and mass flux}
The macroscopic FFD approach is presented without magnetodissipation, i.e.
the magnetization ${\bf M}$ is always parallel to the magnetic field ${\bf
H}$, but with dissipative mass fluxes for the two constituents of the MF.
The analysis will result in an analytical expression for the magnetic part
of the chemical potential and a general expression for the mass flux without
any assumption about the properties of the MF and the temperature
distribution.

The principal structure of the ferrofluid dynamics theory was given in
\cite{mueller01}. It is based firstly on general principles as symmetry considerations
and conservations laws and on irreversible thermodynamics. The second
independent component of which a macroscopic theory is made of is the set
of material-dependent parameters as susceptibilities and transport coefficients.
The latter can be determined by suitable experiments
which are used here to determine transport coefficients for thermodiffusive processes
in magnetic fluids in the presence of magnetic fields.

As usual in theories based on thermodynamical considerations, one starts with
the thermodynamic energy density $u$. It is taken as a function of the entropy
density $s$, the density $\rho^{(1)}$ of the magnetic part of the fluid, the
momentum density ${\bf g}=\rho{\bf v}$, the total density $\rho$, and the
magnetic induction ${\bf B}=\mu_0({\bf M} + {\bf H})$ \cite{mueller01},
\begin{equation}
\label{eq:1}
du = Tds +\tilde\mu_c d\rho^{(1)} +v_i dg_i+\mu^{(2)}d\rho + H_i dB_i \; ,
\end{equation}
where $\tilde\mu_c=\tilde\mu^{(1)} - \tilde\mu^{(2)}$ is the difference
in the chemical potentials of the two constituents. The conservation laws
for the density of the magnetic and nonmagnetic part $\rho^{(2)}$ are
\begin{eqnarray}
\label{eq:2}
\partial_t \rho^{(1)} &=& -\nabla_i \left( \rho^{(1)} v_i -j_i^D\right)\; ,\\
\label{eq:3}
\partial_t \rho^{(2)} &=&-\nabla_i\left( \rho^{(2)} v_i +j_i^D\right)\; ,
\end{eqnarray}
where $j_i^{D(1)}\!=\!-j_i^{D(2)}\!=\!j_i^D$ was used to ensure the conservation
of the total density $\rho\!=\!\rho^{(1)}+\rho^{(2)}\!=\!\phi\rho_{\rm m}
+(1-\phi)
\rho_{\rm cl}$. The density of the magnetic particles (carrier liquid) is
denoted
by $\rho_{\rm m}$ ($\rho_{\rm cl}$) and $\phi$ is the volume fraction of
magnetic particles
in the fluid. The dissipative mass flux ${\bf j}^D$ is proportional
to the gradient of the chemical potential with
$\tilde\mu_c = \tilde\mu_c (\rho , \rho^{(1)}, T, {\bf v}, {\bf H})$
and the temperature gradient \cite{groot_mazur84}. It is assumed that the
magnetic part
of the chemical potential can be separated \cite{landau_lifshitzVIII},
\begin{equation}
\label{eq:4}
\tilde\mu_c = \mu_c (\rho ,\rho^{(1)}, T, {\bf v}) +\mu_c^m(\rho , \rho^{(1)},
T, {\bf v}, {\bf H})\; .
\end{equation}
This assumption guarantees a nonzero chemical potential for ${\bf H}=0$
and is confirmed by calculations for MF with chains, where the magnetic
part contributes additive to the total chemical potential \cite{zubarev02}.
The nonmagnetic part of the chemical potential is given by
$\mu_c\!=\!(k_B T/m_{\rm m})\ln c_1 - (k_{\rm B} T/m_{\rm cl})\ln c_2$,
where $m_{\rm m}$ ($m_{\rm cl}$) is the mass of a magnetic (carrier liquid)
particle \cite{groot_mazur84}.

The experiments \cite{voelker02_phdBook,voelker03,voelker03_prl}
show that {\it any} nonzero strength of the magnetic field
influences the thermodiffusive processes. Thus the general ansatz for the
dissipative mass flux is (following the notation in \cite{mueller01})
\begin{equation}
\label{eq:5}
j_i^D = \xi_1\nabla_i T +\xi \nabla_i \tilde\mu_c +\xi_\parallel M_i M_j
\nabla_j \tilde\mu_c
+ \xi_\times \varepsilon_{ijk}M_j \nabla_k \tilde\mu_c\; .
\end{equation}
Whereas the first two terms characterize isotropic mass fluxes caused by
gradients in the temperature or in the chemical potential, the last two terms
describe anisotropic mass fluxes, namely parallel and perpendicular to the
direction of ${\bf M}$. The last term corresponds to that one in the analogous
ansatz for the heat flux, where the phenomenon is called transversal
Righi-Leduc effect \cite{groot_mazur84} since the primary current is perpendicular
to the produced effect.

It was emphasized in \cite{mueller01} that the "proliferation of transport
coefficients", i.e. $\xi \rightarrow (\xi , \xi_\parallel , \xi_\times)$,
takes place in the case of strong magnetic fields. But the experiments
show that at least for thermodiffusive processes that general statement
seems not to be true. In the figures in \cite{voelker02_phdBook,voelker03,voelker03_prl}
with respect to the changes of $S_T$ it is evident that small magnetic
fields in the order of less than 50 kA/m are sufficient to generate effects,
where one can clearly distinguish between a parallel or a perpendicular
orientation between temperature gradient and field. Therefore the in
\cite{mueller01} firstly introduced coefficients $\xi_\parallel$ and
$\xi_\times$ are considered here as nonzero for all magnetic field strengths.

With the above given dependences of the chemical potential in Eq.~(\ref{eq:4}),
its gradient is
\begin{equation}
\label{eq:6}
\nabla_i \tilde\mu_c = {\partial \tilde\mu_c \over \partial \rho}\nabla_i \rho
+{\partial\tilde\mu_c \over \partial \rho^{(1)}}\nabla_i \rho^{(1)}
+{\partial\tilde\mu_c \over \partial T}\nabla_i T
+{\partial\tilde\mu_c \over \partial v_j}\nabla_i v_j
+{\partial \mu_c^m \over \partial H_j}\nabla_i H_j\; .
\end{equation}
The first expression in Eq.~(\ref{eq:6}) will become later the term for the
barodiffusion and can be neglected in an incompressible fluid not subjected to
any pressure gradient. For the fourth and fifth term
\cite{bacri95_forced,bacri95_transient} hold
\begin{eqnarray}
\label{eq:7}
{\partial \tilde\mu_c \over \partial v_j} &=& -{\partial (\rho v_j)\over
\partial
\rho^{(1)}}\equiv 0 \; ,\\
\label{eq:8}
{\partial \mu_c^m \over \partial H_j} &=& -\mu_0{\partial\over
\partial\rho^{(1)}}
(H_j + M_j) = - \mu_0{\partial M_j\over \partial\rho^{(1)}}\; .
\end{eqnarray}
The transformation $\tilde u= u-v_j g_j-
H_jB_j$ was made in order to match the dependences of the energy
density and the chemical potential and usage of the fact that derivatives of
quantities are zero which are independent of each other. From the last
equality the analytical result for the magnetic part of the chemical
potential follows,
\begin{equation}
\label{eq:9}
\mu_c^m= -\mu_0\int_0^H{\partial M\over\partial\rho^{(1)}}dH' \; ,
\end{equation}
where $M$ and $H$ denote the absolute value of the magnetic field and the
magnetization. Eq.~(\ref{eq:9}) allows a direct determination of $\mu_c^m$ if the
magnetization $M(H,\rho^{(1)},T)$ is known without any assumption about the properties
of the MF in contrast to \cite{zubarev02,blums_book,bacri95_forced,bacri95_transient}.
According to these references the determination of the chemical potential needs the
knowledge of quantities like the volume concentration of the nanoparticles
\cite{zubarev02,blums_book} or the strength of the magnetodipole
interaction \cite{zubarev02} or the effective field experienced by a single
particle in the MF \cite{blums_book,bacri95_forced,bacri95_transient}.
Compared with the effort
to evaluate these microscopic details, the advantage of the macroscopic approach
of the FFD is apparent. A measurement of the magnetization as function of the
magnetic field and the density is sufficient to determine the chemical potential
for any magnetic fluid.

Inserting Eq.~(\ref{eq:6}) into Eq.~(\ref{eq:5}) and using Eqs.~(\ref{eq:7},
\ref{eq:8}) an expression for the mass flux results,
\begin{eqnarray}
\nonumber
{{\bf j}^D\over \rho} &=&
 \left( {\xi_1\over \rho} +{\xi\over \rho}{\partial\mu_c \over \partial T}\right)\nabla T
+{\xi\over \rho}{\partial \mu_c^m\over \partial T}\nabla T
+{\partial \tilde\mu_c \over \partial T}\left[ {\xi_\parallel\over \rho}{\bf M}\left({\bf M}\nabla T\right)
+{\xi_\times\over \rho}\left({\bf M}\!\times\! \nabla T \right)\right]\\
\nonumber
&&
+\xi{\partial\mu_c \over  \partial\rho^{(1)}}\nabla c_1
+\xi{\partial \mu_c^m \over  \partial\rho^{(1)}}\nabla c_1
+{\partial \tilde\mu_c \over  \partial\rho^{(1)}}\left[ \xi_\parallel {\bf M}\left({\bf M}\nabla c_1\right)
+\xi_\times \left({\bf M}\!\times\! \nabla c_1 \right)\right]\\
\label{eq:10}
&&
-{\mu_0\over \rho}{\partial M\over \partial\rho^{(1)}} \left[ \xi\nabla H  +\xi_\parallel {\bf M}
 \left({\bf M}\nabla H\right)
+\xi_\times \left({\bf M}\!\times\! \nabla H \right)\right] \; ,
\end{eqnarray}
which is generally valid, independent of the size distribution of the
magnetic particles, concentration inhomogeneities in the suspension
or the form of the temperature gradient. Therefore Eq.~(\ref{eq:10}) is
the generalization of the mass flux given in \cite{blums98}. The concentration
of the magnetic particles $c_1=\rho^{(1)}/\rho$ is defined by means of the
mass fraction of the total density $\rho$ \cite{groot_mazur84}. The first
four terms describe mass flow caused by thermophoresis ($\sim\!\nabla T$), the
second four terms by diffusiophoresis ($\sim\! \nabla c_1$),
and the last three by magnetophoresis ($\sim\! \nabla H$). There are two
unknown transport coefficients, $\xi_\parallel$ and $\xi_\times$, since for
zero magnetic field, Eq.~(\ref{eq:10}) reduces to the classical
result (see Eq.~(227), Chapt. XI in \cite{groot_mazur84})
\begin{eqnarray}
\nonumber
{{\bf j}^D\over \rho} &=& \left( {\xi_1\over \rho} +{\xi\over \rho}
{\partial\mu_c \over \partial T}\right)\nabla T
+\xi{\partial\mu_c \over  \partial\rho^{(1)}}\nabla c_1\\
\label{eq:11}
&=&c_1 c_2 D_T \nabla T + D_c \nabla c_1
\end{eqnarray}
with ($D_T$) $D_c$ the (thermal) diffusion coefficient known for MFs
from previous experiments \cite{lenglet02} and $c_2=1-c_1$. According to the
philosophy of the FFD approach, the determination of the unknown transport
coefficients $\xi_\parallel$ and $\xi_\times$ needs suitable experiments
which were conducted just recently \cite{voelker03_prl}.

\section{Application to experiments and discussion}
According to the experiments for a horizontal layer of MF of thickness $h$
\cite{voelker03_prl}, a horizontally unbounded layer of a dielectric, viscous,
and incompressible MF sandwiched between two perfect conducting plates is
considered. The lower plate is cooled to $T_1$ and the upper one is heated to
$T_2$. The resulting temperature gradient stabilizes the quiescent conductive
state. From the equation of heat conduction,
\begin{equation}
\label{eq:11.1}
{\partial T\over \partial t} = \kappa\Delta T\; ,
\end{equation}
and the boundary conditions,
\begin{equation}
\label{eq:11.2}
T(z=h/2)=T_2 \qquad\qquad  T(z=-h/2)=T_1\; ,
\end{equation}
the temperature profile of the conductive state
\begin{equation}
\label{eq:11.3}
T=T_0+{(T_2-T_1)\over h}z
\end{equation}
follows with $T_0\!=\!(T_1+T_2)/2$ and $\kappa$ denotes the thermal diffusivity.
Since the plates are impenetrable, the diffusion equation,
\begin{equation}
\label{eq:11.4}
{\partial c_1\over \partial t} = \diver \left( {{\bf j}^D\over \rho}\right) \; ,
\end{equation}
has to be supplemented with the boundary condition
\begin{equation}
\label{eq:11.5}
j_z^D(z=\pm h/2)=0.
\end{equation}
Rearranging this boundary condition with the help of Eq.~(\ref{eq:11}),
\begin{equation}
\label{eq:12}
{-h \over c_1 c_2 (T_2 -T_1)}{\partial c_1\over \partial z}\biggr|_{z=\pm h/2}
={D_T \over D_c} = S_T\; ,
\end{equation}
the Soret coefficient in the zero field case can be expressed. In the same
way the global Soret coefficient, measured in \cite{voelker03_prl},
in the presence of a magnetic field can be determined by using Eq.~(\ref{eq:10}).

If a spatially homogeneous static magnetic field is applied perpendicular to
the layer, the resulting magnetic field gradient inside the fluid is parallel to
the temperature gradient. Therefore this setup is called parallel and is
analyzed first.

Taking diffusion processes into account, the magnetization in the fluid can be
written in the form
\begin{equation}
\label{eq:13}
{\bf M}=\left[ M_0 +\chi (H-H_0)-K(T-T_0)+{\partial M\over \partial \phi}(\phi
-\phi_0)\right]{\bf e}_z\; ,
\end{equation}
where $M_0=M_0(H_0, T_0, \phi_0)$ is the reference magnetization belonging to
the reference values $H_0$, $T_0$, and $\phi_0$ for the magnetic field, the
temperature, and the volume fraction. Extending the expressions given in
\cite{auernhammer00}, magnetization and magnetic field for the conductive state are
\begin{eqnarray}
\label{eq:14}
M &=& M_0 + {K(T_1-T_2)\over h\,(1+\chi)}z + N (c_1-c_{1,0})\; ,\\
\label{eq:15}
H &=& H_0 - {K(T_1-T_2)\over h\,(1+\chi)}z - N (c_1-c_{1,0})\; ,
\end{eqnarray}
with the susceptibility $\chi = \partial M/\partial H$, the
pyromagnetic coefficient $K =-\partial M/\partial T$, the
densomagnetic coefficient $N=\partial M/\partial c_1 =
(\rho/\rho_m)(\partial M/\partial \phi)$, and $c_{1,0}=c_1(T_0)$.
Inserting Eqs.~(\ref{eq:14}, \ref{eq:15}) into Eq.~(\ref{eq:10}) and rearranging
the boundary condition~(\ref{eq:11.5}) in the same manner as in the zero field case,
the global Soret coefficient in the parallel setup reads:
\begin{equation}
\label{eq:16}
S_T^\parallel = {S_T +\displaystyle{{1\over c_{1,0}\, c_{2,0}\, \rho}\left[
{\xi\over D_c}{\partial\mu_c^m\over \partial T}+{\xi_\parallel\over D_c}
{\partial\mu_c\over \partial T} M^2-\mu_0{\partial M\over \partial\rho^{(1)}}
\left( {\xi\over D_c}+{\xi_\parallel\over D_c}M^2\right){K\over (1+\chi)}
\right] }\over \displaystyle{1+{\xi\over D_c}{\partial\mu_c^m\over
\partial\rho^{(1)}} + {\xi_\parallel\over D_c}{\partial\mu_c\over
\partial\rho^{(1)}}M^2 + {\mu_0\over \rho}{\partial M\over \partial\rho^{(1)}}
\left( {\xi\over D_c}+{\xi_\parallel\over D_c}M^2 \right) N
}}\; .
\end{equation}

\begin{figure}[htbp]
  \begin{center}
    \includegraphics[width=8.6cm]{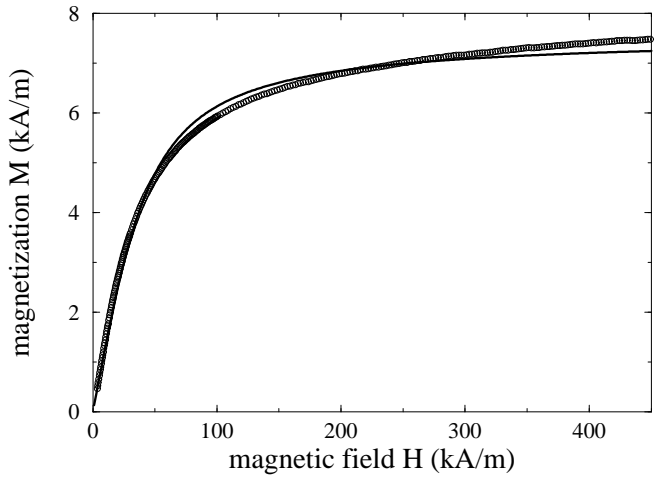}
    \caption{Experimental data ($\circ$) from Fig.~57 in
    \cite{voelker02_phdBook} and theory
    (solid line) for the magnetization at room temperature $T=293$ K.
    The details of the used Langevin function are given in the text.}
    \label{fig:magnetization}
  \end{center}
\end{figure}

Knowing $M(H,\rho^{(1)},T)$ in analytical form allows one to
calculate $\mu_c^m$ and its derivatives. The measured magnetization curve
(from Fig.~57 in \cite{voelker02_phdBook}) could be nicely fitted
with $M = M_b \lambda_\phi \phi L(\lambda_d \alpha)$, where
$L(\lambda_d \alpha)=\coth (\lambda_d \alpha)-1/(\lambda_d \alpha)$
is the Langevin function, $\alpha = \mu_0 m H/(k_{\rm B} T)$ the Langevin
parameter, $m = M_b\pi d^3/6$ the magnetic moment of a particle, and
$k_{\rm B}$ the Boltzmann constant. $\lambda_d$ and
$\lambda_\phi$ are two geometrical fit parameters. They reflect
small deviations from the volume fraction $\phi\! =\! 0.2$
and the $\delta$-shaped size distribution (Fig.~59 in
\cite{voelker02_phdBook}).
Using $\lambda_d= 0.99$, $\lambda_\phi = 0.84$ and the material
data $M_b =450$ kA/m (magnetization of the magnetic bulk solid),
$d =9$ nm, $\rho_m = 5.15$ g/cm$^3$ from \cite{voelker02_phdBook},
the solid line in Fig.~\ref{fig:magnetization} shows a very good agreement
with the measured magnetization ($\circ$). Considering the chosen values for
$\lambda_d$ and $\lambda_\phi$, only the volume fraction $\phi$ had to be adjusted
to the measured data. Variations in $\phi$ are likely caused by
a nonmagnetic surface layer of the nanoparticles \cite{blums97} and its
solvability in the carrier liquid. According to the statement at the
beginning of the paragraph, one has
\begin{equation}
\label{eq:16.1}
\mu_c^m = {\lambda_\phi\over \lambda_d}{M_b\, k_{\rm B} T\over 2\rho_{\rm m}
m}\left\{ \ln\left[ \coth^2(\lambda_d \alpha)-1\right] + 2\ln (\lambda_d \alpha)\right\}\; ,
\end{equation}
from which one can calculate the derivatives with respect to $T$ and $\rho^{(1)}=\phi\rho_m$.

\begin{figure}[htbp]
  \begin{center}
    \includegraphics[width=8.6cm]{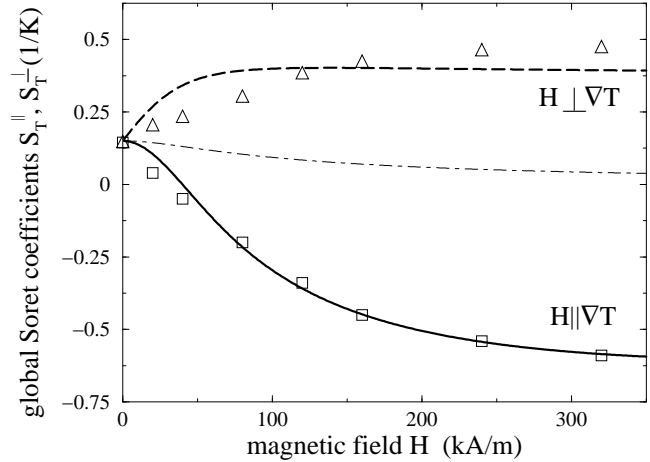}
    \caption{Global Soret coefficients  $S_T^\parallel$ and $S_T^\perp$
    against the magnetic field strength for the
    parallel ($H\parallel\nabla T$) and perpendicular setup ($H\!\!\perp\!\!\nabla T$).
    The solid line shows the best fit of $S_T^\parallel$ [see Eq.~(\ref{eq:16})]
    with $\xi/D_c\!=\!8.2$ kg$\,$s$^2/$m$^5$ and
    $\xi_\parallel/ D_c\!=\!-1.41\cdot 10^{-7}$ kg$\,$s$^2/($m$^3\,$A$^2)$ to
    the experimental data ($\Box$). The dot-dashed line displays $S_T^\parallel$
    for the same parameters but  $\xi_\parallel/ D_c\!=\!0$. The dashed line
    indicates the best fit of $S_T^\perp$ [see Eq.~(\ref{eq:19})] with
    $\xi/D_c\!=\!8.2$ kg$\,$s$^2/$m$^5$ and $F\! =\!3.75\cdot 10^{-2}$ kg$\,$s$^2/($m$^5$A)
    to the experimental data ($\triangle$). For $F$  and all other values see
    text.}
    \label{fig:ex_theo}
  \end{center}
\end{figure}

With the pyromagnetic coefficient $K$ taken from Fig.~4 in \cite{voelker02_phdBook},
it remains the four unknowns $S_T$, $D_c$, $\xi$, and $\xi_\parallel$ in
Eq.~(\ref{eq:16}). To fit $S_T^\parallel$ to the experiment, the combined quantities $\xi/D_c$
and $\xi_\parallel/ D_c$ are used as fit parameters, since $S_T\!=\!0.15$
K$^{-1}$ was measured in the zero field case \cite{voelker03_prl}
but not $D_c$. The solid line in
Fig.~\ref{fig:ex_theo} gives the best two parameter fit, yielding
$\xi/D_c\!=\!8.2$ kg$\,$s$^2/$m$^5$ and
$\xi_\parallel/ D_c\!=\!-1.41\cdot 10^{-7}$ kg$\,$s$^2/($m$^3\,$A$^2)$. The
difference
in the absolute values of about $8$ orders of magnitude is not surprisingly,
since one would assume such a relation according to the argument that
anisotropic fluxes in the mass flux~(\ref{eq:5}) are relevant only for strong
fields \cite{mueller01}. Inspecting Eq.~(\ref{eq:16}) closer, it is
revealed that $\xi_\parallel/ D_c$ is
multiplied by $M^2$ which gives already for small magnetic fields a factor of
$\sim\!\!10^6$. The
two other terms are not so relevant
because $0\!\leq\! \partial\mu_c/\partial T \!\leq\! 0.016\,$J/(K$\,$kg) and
$0\leq  \partial\mu_c/\partial\rho^{(1)}\leq 0.35\,$Jm$^3$/kg$^2$ for $0\leq H
\leq
350\,$kA/m. To underline the relevance of $\xi_\parallel/ D_c$ even for
small fields, the dot-dashed line in Fig.~\ref{fig:ex_theo} displays
$S_T^\parallel$ for $\xi_\parallel/ D_c\!=\!0$ and all other parameters
as before. Now the theoretical curve misses the measured data ($\Box$)
clearly. Taking a typical value for the diffusion coefficient,
$D_c\sim 10^{-11}$ m$^2/$s \cite{lenglet02}, the new transport
coefficient can be estimated to
$\xi_\parallel\sim -10^{-18}$ kg$\,$s$/($m$\,$A$^2)$
for the MF in \cite{voelker03_prl,voelker02_phdBook}.
Thus those experiments deliver the necessary input for determining
the material-dependent transport coefficients which are a priori unknown
in a macroscopic theory as the FFD. Another example for the experimental
determination of diffusion and thermodiffusion coefficients is presented
in \cite{lenglet02}, whereas in \cite{bringuier03} these coefficients were
calculated on the basis of a microscopic theory.

In contrast to the parallel setup, in the perpendicular setup the spatially
homogeneous static magnetic field is applied perpendicular to the temperature
gradient, i.e. the magnetic field is parallel to the layer.
The diffusion equation gets now the form
\begin{equation}
\label{eq:17}
{\partial c_1\over \partial t} ={\partial \tilde\mu\over\partial \rho^{(1)}}
\left[ (\xi+\xi_\parallel M^2){\partial^2 c_1\over \partial x^2}
+\xi\left({\partial^2 c_1\over \partial y^2}+{\partial^2 c_1\over \partial z^2}
\right)\right] \; .
\end{equation}
The boundary condition for the $z$-component of the mass flux yields
\begin{equation}
\label{eq:18}
{\partial c_1\over \partial z} = -{\xi_\perp M\over \xi}{\partial c_1\over \partial y}
-{\displaystyle{c_1\, c_2\, D_T + {\xi\over \rho}{\partial\mu_c^m\over \partial T}}
 \over\displaystyle{D_c +\xi{\partial\mu_c^m\over \partial \rho^{(1)}}}}
{\rm ~at~}z=\pm h/2\; .
\end{equation}
Since no analytical solution for that boundary value problem is known,
the following coarse approximation is made: $(\partial c_1/\partial
y)_{z=\pm h/2}$ shall be a constant $C$ for all $H$-values tested here.
The global Soret coefficient in the perpendicular setup can be approximated than by
\begin{equation}
\label{eq:19}
S_T^\perp = {h\over (T_2-T_1)\,c_1\, c_2}F{M\over \displaystyle{\xi\over D_c}}
+{\displaystyle{ S_T +{\xi\over D_c}{1\over c_1\, c_2\,\rho}{\partial\mu_c^m\over \partial T}}
 \over\displaystyle{ 1+ {\xi\over D_c}{\partial\mu_c^m\over \partial
\rho^{(1)}}}}\; ,
\end{equation}
where $F\!=\!(\xi_\perp C)/D_c$ will be used as the only fit parameter since
$\xi/D_c$ was determined in the parallel setup. With $T_2-T_1\!=\!1$ K,
$\phi\!=\!0.2$, and $h\!=\!1$
mm \cite{voelker03_prl}, the best fit yields $F\! =\!3.75\cdot 10^{-2}$ kg$\,$s$^2/($m$^5$A).
The inferior match with the experimental data (see $\triangle$ and dashed line
in Fig.~\ref{fig:ex_theo}) in comparison
with the parallel setup is due to the approximation that
$(\partial c_1/\partial y)_{z=\pm h/2}$ is constant. In the real system it
will depend on the magnetic field since the solution for $c_1$ depends on the
magnetic field.

\section{conclusion}
The ferrofluid dynamics theory is applied to thermodiffusive
problems in magnetic fluids in the presence of magnetic fields, where the
MF is considered as a binary mixture. In the frame work of this
theory the chemical potential could be determined analytically.
Also a general expression for the mass flux is given which is
independent of the fluid properties, temperature distribution and
assumptions about the concentration of the nanoparticles.
Applying these results to the experiments \cite{voelker03_prl}, their
data could be interpreted better (see Fig.~\ref{fig:ex_theo}) than
with the previous theory \cite{blums98} which gave values
about three orders of magnitude too small. Three transport coefficients,
which are inherent parts the macroscopic
ferrofluid dynamics theory \cite{mueller01}, had to be
used to fit this theory with the only sets of experiments available at
present. In general, it is shown that for thermodiffusive problems in
magnetic fluids, i.e. in colloidal suspensions sensitive to external fields,
anisotropic mass fluxes are relevant and no small contributions for any
nonzero strengths of the magnetic field. To elucidate this insight, more
well designed experiments and further theoretical analyses are needed to
improve the knowledge about thermodiffusive processes in magnetic fluids.

\section{Acknowledgement}
The author would like to thanks Thomas V\"olker for providing the
experimental data and Konstantin Morozov as well as Stefan Odenbach
for stimulating discussions.

\bibliography{h:/tex/bib/mf_general,h:/tex/bib/mf_conv}

\end{document}